\documentclass[showpacs,floatfix,preprint,nofootinbib]{revtex4}
\usepackage{bm}

\let\al=\alpha

\let\d=\delta

\let\om=\omega

\begin{document}
\title{ Is violation of Newton's second law possible?}
\author{A.~Yu.~Ignatiev}
\email{a.ignatiev@ritp.org}
 \affiliation {{\em Theoretical Physics Research Institute,
     Melbourne 3163,}\\
    {\em   Australia.}}

\def\be{\begin{equation}}
\def\ee{\end{equation}}
\def\bea{\begin{eqnarray}}
\def\eea{\end{eqnarray}}
\newcommand{\nn}{\nonumber \\}

\begin{abstract}
Astrophysical observations (usually explained by dark matter) suggest that classical mechanics could break down when the acceleration becomes extremely small 
(the approach known as modified Newtonian dynamics, or MOND). 
I present the first analysis of MOND manifestations in terrestrial (rather than astrophysical) settings.
A new effect is reported:  around each equinox date,
2 spots emerge on the Earth where static bodies experience spontaneous acceleration due to the possible violation of  Newton's second law. 
Preliminary estimates indicate that an experimental search for this effect can be feasible.
\end{abstract}
\maketitle
\section{Introduction}

The work described here is motivated by a long-standing  puzzle of astrophysics:  why does
 matter rotate around the centers of
galaxies {\em faster} than expected? This could be due 
to
(1)
dark matter around the galaxies,
(2) modification of the 
Newtonian gravitational law, or 
(3) corrections to the second Newton's law. 
Possibilities (2) and (3) were proposed by M.~Milgrom in 1983 \cite{m0}; they are known collectively as modified Newtonian dynamics (MOND).
While the dark matter studies are
 more common,
 interest in the MOND alternative  is rapidly growing at the moment \cite{m,s}.
 Several recent reviews of MOND's successes and  
challenges are available \cite{rev}. 

One may wonder if there is any point in questioning Newtonian mechanics which has been with us for over 3 centuries and has never failed (within its area of applicability). The answer  is that the MOND effects could only take place
in a very special regime: the accelerations must be unusually small, of the order of $a_0\simeq2\times10^{-10}\; m\,s^{-2}$. 
The following modification of the 2nd Newton's law
 would fit the astrophysical data:
$\mathbf{F}=m\mathbf{a}\mu(a/a_0)$, where $\mu$ is a function satisfying the two conditions: $\mu(a/a_0) \rightarrow 1$ at $a\gg a_0$ and $\mu(a/a_0) \rightarrow 0$ at $a\ll a_0$.
Such small accelerations very rarely occur under ordinary (i.e., non-astrophysical) circumstances, and thus possible MOND effects could  easily have gone unnoticed.
However, should MOND turn out to be correct, then the foundations of physics, including classical mechanics and general relativity, would have to be revised.

This explains why  ground-based laboratory tests of MOND  are of vital importance not only for astrophysics and cosmology, but also for modern physics as a whole. 
Yet  due to perceived difficulties such tests have never been attempted, or even seriously discussed.

In this paper I show that this perception 
can be overcome. It turns out  possible to predict exactly when,  where, and under what conditions the MOND effects would manifest themselves on the Earth.
The existing experimental accuracy appears to be close or better than the precision required for the MOND-testing purposes.
  As a result, several different experimental set-ups can be imagined. I also formulate the most general 
conditions that {\em any} MOND-testing set-up should satisfy.

\section{Acceleration: `absolute' or  `relative'?}
First I  emphasize that to obtain laboratory-testable predictions,
  MOND needs to be formulated not only in inertial reference systems, but also in non-inertial systems as well. (In the MOND context all laboratory reference systems should be considered as  non-inertial.) Because  the dynamical law is modified depending on the acceleration, the transition between inertial and non-inertial systems in MOND becomes less straightforward than in the conventional mechanics. 

Of particular interest are  transformation properties
of  $a_0$.  
 Logically, at least 2 options could be imagined. First, one can assume that the fundamental acceleration that determines the onset of the MOND regime equals $a_0$ only in the inertial reference systems.
Second, it could be assume that $a_0$
 is invariant under transformations from inertial to non-inertial systems. One would expect that these 2 types of theories would lead to drastically different experimental predictions.

For instance, the first  type of theory requires that the MOND regime is reached as soon as the test body moves with a tiny acceleration $\alt a_0$ \emph{with respect to the Galactic reference frame}.

On the other hand,  the second type  of theory implies that in order to reach the MOND regime, we should try to ensure that the test body moves with a tiny acceleration $\alt a_0$ \emph{with respect to the laboratory reference frame}. 

However, it  can be shown that 
the second version (invariant acceleration $a_0$) is not self-consistent. The reason is that the invariance of $a_0$ is inconsistent with the kinematical rules of acceleration addition. In what follows, only the first version will be considered.

\section{ The general condition of entering the MOND regime}
We will now analyze what conditions must be realized in order to obtain  the MOND effect for  test bodies moving in the ground-based laboratory.

This question is easy to answer in the inertial system $S_0$. (It is the system with the origin in the centre-of-mass of our Galaxy and the axes pointing to certain far-away quasars). In this system, we should ensure that the test body moves with a tiny acceleration ${\bf a}_{gal}$ with respect to $S_0$:
\begin{equation}
\label{0}
\mathbf{a}_{gal}\approx 0.
\end{equation}
Throughout the paper, the $\approx$ sign will mean that the difference between the left-hand side and the right-hand side of an equation is much less than the characteristic MOND acceleration $a_0$.

Next, we are going to the laboratory system with the help of 
\begin{equation}
\label{0a}
\mathbf{a}_{gal}\approx\mathbf{a}_{lab} +\mathbf{a}_1(t)+{\bm\om}\times(\bm{\om}\times(\mathbf{r}+\mathbf{r}_1))+2 \bm{\om}\times\mathbf{v}+\mathbf{a}_2,
\end{equation}
where $\mathbf{a}_1$ is the acceleration of the Earth's centre with respect to the heliocentric reference frame,
$\bm{\om}$ is the angular velocity of the Earth's rotation, $\mathbf{a}_2$ is the Sun's acceleration with respect to $S_0$;
 $\mathbf{r},\; \mathbf{v}=\dot{\mathbf{r}}$, and $\mathbf{a}_{lab}=\ddot{\mathbf{r}}$ are the position, velocity, and acceleration of the test body with respect to the laboratory reference frame;  $\mathbf{r}_1$ is the position vector of the origin of the lab frame with respect to the terrestrial frame with the origin at the Earth's centre \footnote{As practical, high-precision realizations of these intermediate frames one can take the International Celestial Reference System (ICRS) \cite{icrs} and the International Terrestrial Reference System (ITRS) \cite{itrs}.}. A number of terms have not been written out in Eq.~(\ref{0a}) on account of their smallness. They include terms due to: the Coriolis acceleration of the Sun, the length-of-day variation, precession and nutation of the Earth's rotation axis, polar motion and Chandler's wobble.
From Eq.~(\ref{0}) and Eq.~(\ref{0a}) we obtain the necessary and sufficient condition for realisation of the MOND regime in the laboratory:
\begin{equation}
\label{2}
\mathbf{a}_{lab}\approx -\mathbf{a}_1(t)-\bm{\om}\times(\bm{\om}\times(\mathbf{r}+\mathbf{r}_1))-2 \bm{\om}\times\mathbf{v}-\mathbf{a}_2.
\end{equation} 

\section{The `SHLEM' effect} 
The simplest set-up for implementing this condition would be to have a test body that is at rest in the laboratory frame. Can we test MOND in this way? If we put $\mathbf{v} = 0, \; \mathbf{a}_{lab}=0$, and (without loss of generality) $\mathbf{r}=0$,  the above equation becomes
\begin{equation}
\label{2.0}
 \mathbf{a}_s(t)+\bm{\om}\times(\bm{\om}\times \mathbf{r}_1)\approx 0,
\end{equation}
where I  have introduced $ \mathbf{a}_s =  \mathbf{a}_1+  \mathbf{a}_2$ for convenience. We note that this equation has no solutions unless $\mathbf{a}_s$ is orthogonal to $\bm{\om}$, so we must first look for those instants $t_p$ when
\begin{equation}
\label{55}
a_{s\parallel}(t_p) \approx 0 \;  or \;  a_{s\parallel}(t_p)| \ll a_0,
\end{equation}
where $a_{s\parallel}=(\mathbf{a}_s\bm{\om})/\om$.
A continuity argument shows that this equation has at least 2 solutions during each year. Indeed, at the instant of a (nothern) summer solstice $a_{s\parallel} > 0$ whereas at the instant of  a  winter solstice $a_{s\parallel} < 0$. Therefore, there must be at least one instant during autumn and one instant during spring when $a_{s\parallel} = 0$ {\em exactly}. Neglecting the effects due to the Moon and planets, these instants would coincide exactly with the autumnal and vernal equinoxes. In reality, the instants will be shifted from the equinoxes. However, the above `existence theorem' guarantees that these instants $t_p$ can be found with astronomical precision through a straightforward but time-consuming procedure using the lunar and planetary ephemerides. In addition, one can show that the off-equinox shift, in any case, should be less than a few days.

Another estimate shows that due to the Earth's orbital motion,  Eq.~(\ref{55}) will only stay valid for     
the time interval of the order of $\d t\sim (a_0/a_s)(4\epsilon/T)^{-1}\sim1\; s$ where $\epsilon=23^o27'=0.41$, $T=1\; yr$.\footnote{Strictly speaking, one should also consider the analogous interval $\d t'$ due to the lunar orbital motion and then pick up the shorter of the two. However, $\d t'\sim (a_0/a')(4\epsilon'/T')^{-1}$ turns out to be larger than $1\;s$ due to the small ratio $a'/a_s \simeq 1/180$ and, thus, this point can be ignored.}
Once  $t_p$ is found and plugged into Eq.~(\ref{2.0}), the corresponding solution for the laboratory location is $\mathbf{r}_{1\perp}=  \mathbf{a}_s(t_p)/\om^2$.  This  key relation allows us to find both the latitude and the longitude of the right spot. If we again ignore the lunar and planetary effects, the relevant magnitude is $|\mathbf{a}_s(t_p)| \simeq 0.00593  \; m\,s^{-2}$ which gives the required latitude  $\phi\simeq\pm 79^o50'$. As for the longitude, it would generally vary from year to year. For instance, on the autumnal equinox of September 22, 2008 these spots would be at $56^o$ West longitude---one in Greenland, ($79^o50'$ North latitude), another in Antarctica ($79^o50'$ South latitude). The account of lunar perturbation can significantly change the longitude, but the latitude prediction is much more robust: it would not change by more than $\sim 6'$, or $10\; km$.

To emphasize these conditions, I will use the acronym ``SHLEM'' (Static High-Latitude Equinox Modified inertia).
The signature of the SHLEM effect would be a spontaneous displacement of the test body occurring exactly at the instant $t_p$ defined by Eq.~(\ref{55}).
The displacement amplitude would be  of the order of $a_0 \tau^2/2 \sim 0.2\times 10^{-16}\; m$, with the effective dynamic-violation time\footnote{The interval $\tau$ is determined by the Earth's rotation around its axis.  Fortunately, this interval is longer than the characteristic timescale of a LIGO-type interferometer which is set by the ``round-trip time'' $2L/c\simeq3\times10^{-5}\; s$ where $L = 4\; km$ is the interferometer arm length.} $\tau\sim a_0/(\omega a_s)\simeq 0.5\;ms$. This can be compared with the current sensitivity of gravitational wave detectors (such as LIGO, VIRGO, GEO 600, TAMA 300, AIGO and others): about $10^{-18}\;m$ (LIGO) or $3\times10^ {-21}\;m$ (MiniGRAIL, under construction).  Thus it appears  that the use of a similar type of experimental set-up could be an interesting opportunity. Note that the exact prediction of the time of the event will further increase the chances of separating the  SHLEM signal from the noise (and also from true gravitational waves).

Provided that the above three conditions are met, due to the Earth's curvature the SHLEM effect would only be significant in a space box
with the following dimensions:
about $2R_Ea_0\cos{\phi}/a_s\simeq7\;cm$ in the East-West direction and about $2R_Ea_0/a_s\simeq40\;cm$ in the North-South and vertical directions ($R_E$ is the Earth radius). If the laser-interferometer type of gravitational detector is used, then the interferometer's mirror should be placed in such a way as to maximize the overlap between that box and the mirror. 
In particular, the orientation of the mirror is important: for instance, if a thin mirror has a diameter of $25\;cm$ (same as in LIGO) then it should face either East or West. Similar considerations apply to the choice of position and orientation of the detector in the case of low-temperature resonant bar detectors such as AURIGA, NAUTILUS, and ALTAIR (Italy),  EXPLORER (CERN), ALLEGRO (USA), and NIOBE (Western Australia) 
as well as the spherical cryogenic detectors under construction, such as MiniGRAIL, GRAVITON, and TIGA.

The resonant detectors have the advantage of  being more easily transportable. On the other hand, their spectral sensitivity is more narrow than that of the interferometer detectors which could result in some reduction of the SHLEM signal.

In addition to gravitational antennas, one can also think of the torsion balance methods whose existing sensitivity is $\sim10^{-15}\;m\,s^{-2}$ \cite{ad}.  A new design with a better sensitivity
is proposed in Ref.~\cite{h}.

The techniques developed recently for short-range tests of gravity \cite{adrev} can also be of interest in the present context. Indeed, to probe into accelerations of the order of $a_0$ one needs to place 2 masses of the order of $1\;g$ at a distace of the order of a few centimeters.

It would be interesting to consider if other classic gravity experiments---the equivalence principle tests, the fifth force searches etc. (see, e.g., \cite{f}) ---could be adapted for the purposes of searches for the SHLEM effect.

\section{The Coriolis-centrifugal cancellation (`CCC') set-up}
After obtaining the static solution, the next logical step would be to find a stationary (i.e., $\mathbf{v} = const$) solution of Eq.~(\ref{2}). 
Note that the existence of such solution is by no means guaranteed, and indeed we will see that such solution can only be found as an approximation. The physical idea here is the cancellation between the Coriolis and the centrifugal inertial forces which will be referred to as the CCC set-up.

In the stationary case we have ${\bf \ddot r }=0$ and therefore our Eq.~(\ref{2}) takes the form
\begin{equation}
\label{2a}
 \mathbf{a}_s  + \bm{\om}\times(\bm{\om}\times(\mathbf{r}+\mathbf{r}_1)) + 2 \bm{\om}\times\mathbf{v} \approx 0.
\end{equation}
As in the static case, this equation has no solutions unless the orthogonality relation, Eq.~(\ref{55})  holds. Consequently, the above discussion regarding the `orthogonality' instants $t_p$ and the `validity interval' $\d t$ remains in force for the present case as well. 

Now, plugging $\mathbf{r} = \mathbf{r}_0 + \mathbf{v} t$ into Eq.~(\ref{2a}), we obtain
\begin{equation}
\label{2b}
 \mathbf{a}_s  + \bm{\om}\times(\bm{\om}\times(\mathbf{r}_0 +\mathbf{r}_1)) + 2 \bm{\om}\times\mathbf{v}  + \bm{\om}\times(\bm{\om}\times\mathbf{v}) t \approx 0.
\end{equation}
We note that the last term is time-dependent while all other terms do not depend on time \footnote{The time dependence of the first term can be ignored within the `validity interval' $\d t$.}.
Thus the only way to get a solution is to require that the last term is much less than $a_0$. Introducing for convenience a new variable $\mathbf{x} = \bm{\om} \times  \mathbf{v}$, we can write this condition as
\begin{equation}
\label{4}
|\bm{\om} \times  \mathbf{x}| t \ll a_0  \; \; or \;  \;  \mathbf{v}_\perp t  \ll a_0/\om^2 \simeq 4 \; cm,
\end{equation}
where $t$ is the effective duration of the experiment and $\mathbf{v}_\perp $ is the component of $\mathbf{v}$ orthogonal to the Earth  spin.

Assuming that this condition is satisfied and introducing $\mathbf{b} = \bm{\om}  \times (\bm{\om} \times  (\mathbf{r}_1 + \mathbf{r}_0))$ we can now rewrite  Eq.~(\ref{2b})  as follows:
\begin{equation}
\label{2d}
\mathbf{x} \approx - (\mathbf{b} + \mathbf{a}_s)/2.
\end{equation}
The solution is:
\begin{equation}
\label{2i}
\mathbf{v}_\perp \approx \frac{\bm{\om} \times  [ \mathbf{b} + \mathbf{a}_{s\perp}(t_p)]}{2 \om^2},
\end{equation}
where $\mathbf{a}_{s\perp} =\mathbf{a}_s-(\mathbf{a}_s  \bm{\om}) \bm{\om}/\om^2$.
Thus $\mathbf{v}_\perp$ depends  both on the geographic coordinates of the lab and on the orthogonality instant $t_p$.
In summary, Eqs. (\ref{4}, \ref{55}, \ref{2i}) together give the necessary and sufficient conditions for realizing the CCC set-up, i.e., that the motion with the constant velocity given by Eq.~(\ref{2i}) will satisfy Eq.~(\ref{2}).
 Thus the problem  of finding the constant-velocity solution is solved. In contrast with the static case, this solution does not put restrictions on the laboratory location.

\section{Accuracy: existing versus required}

Let us see if the available accuracy of the quantities involved in Eqs.~(\ref{0a}, \ref{2}) is sufficient for our purposes.

For the solar acceleration $\mathbf{a}_2$ one can find from the existing data \cite{a} that 
$|\mathbf{a}_2| = (2.4 \pm 0.3)\times 10^{-10}\; m\,s^{-2}$.
Thus the existing uncertainty in $\mathbf{a}_2$ is about 
15\% of $a_0$. (Note that the angular coordinates of the galactic centre are known so well---within 1 milliarcsec \cite{a}---that  the angular uncertainty in $\mathbf{a}_2$ can be completely ignored.)

The accuracy of Earth's centripetal acceleration $\mathbf{a}_1$ is controlled by the precision $\d k$ in determination  \cite{a} of the Gauss' gravitational constant $k$; presently $\d k/k \simeq 10^{-9}$. Because $|\mathbf{a}_1| \simeq 0.006 \; m\,s^{-2}$, we conclude that the uncertainty in $\mathbf{a}_1$ is about
 3\% of $a_0$. 

The angular velocity of Earth's rotation $\bm{\om}$ is monitored  by the International Earth Rotation and Reference Systems Service (IERS \cite{fr,usno}) with high precision: 
$\d \om/\om \simeq 10^{-12}$. The positions on the Earth's surface, relative to the Earth's centre, can be measured up to $\simeq 1\; mm$ (owing to the International Terrestrial Reference System (ITRS) Product Centre of the IERS  \cite{itrs}). This means that the magnitude of the centrifugal acceleration is known up to about 
5\% of $a_0$ [this figure corresponds to the lab located on the equator; for a non-zero latitutude $\phi$ the accuracy must be multiplied by $(\cos \phi )^{-1}$].

The analysis of the Coriolis term in Eqs.~(\ref{0a},\ref{2}) leads to two sorts of constraints. First, the velocity $\mathbf{v}$ must not be so great that the length-of-the-day uncertainty $\d \om$ would lead to the uncertainty of the Coriolis term of the order of $a_0$.  This condition yields $v \alt (1.4 \times 10^6)/\sin{\al}  \;  m\,s^{-2}$,
where $\al$ is the angle between $\mathbf{v}$ and $\bm{\om}$.
Second, the accuracy of velocity measurement $\d v$ must be such that the corresponding uncertainty in the Coriolis term would be $\ll a_0$. It follows that
$\d v \ll (1.4 \times 10^{-6}/\sin \al) \; m\,s^{-2}.$
This constraint can be rewritten as an upper limit on the velocity $v$ provided the accuracies of the time and length measurements are given. Indeed, suppose that time can be measured with the accuracy of $\d t$ and that the accuracy of length measurement is such that its contribution does not exceed  the contribution of time uncertainty (this assumption seems reasonable because, by definition, the speed of light is known exactly).  Then the resulting upper bound on the velocity can be found as
\begin{equation}
v \ll \left(\frac{10^6}{\sqrt{\sin{\al}}}\right) \sqrt \frac{l}{100\; m}\left(\frac{1}{\d t/10^{-14}\, s}\right) \;m\,s^{-2}.
\end{equation}
Here, $l \sim vt$ is the characteristic distance involved in the experiment. Note that today's  best atomic clock---the mercury clock of the National Institute of Standards and Technology \cite{nist}---has  the accuracy of  about $\d t \simeq 10^{-16} \; s$. 
We conclude that the necessary ingredients 
appear to be known with the precision that is close to the accuracy required by the experiment.

\section{The general solution}
In addition to the particular solutions of Eq.~(\ref{2})---static and stationary---described above, we can also find the general solution of that equation in an analytical form \cite{ign}. This solution requires knowledge of the functions $\mathbf{a}_1(t)$ and $\mathbf{a}_2(t)$, which is provided by astronomical observations. 
To obtain a solution in a more manageable form, I adopt the following simplified model: (1) Acceleration $\mathbf{a}_2(t)$  is ignored (in other words, the heliocentric reference frame is assumed to be inertial). (2) Acceleration $\mathbf{a}_1(t)$ is taken as a harmonic oscillation with the frequency $\om_1 = 2\pi /(1\; yr)$ (i.e., the eccentricity of the Earth's orbit and the Moon's effect are neglected). (3) The direction of $\bm{\om}$ (taken as $z$-axis) is assumed to be orthogonal to the Earth's orbital plane. Then the general solution of Eq.~(\ref{2}) is
\begin{eqnarray}
\nonumber
\ddot x&\approx&  (x_1+x_0) \om^2+2 v_{0y}\om - R\om_1^2 + y_1\om^3t - 3v_{0x}\om^2 t - x_1 \om^4 t^2,\\\label{9}
\ddot y&\approx&  (y_1+y_0) \om^2 - 2 v_{0x}\om  - 2 x_1\om^3t - 3 v_{0y}\om^2 t + v_{0x}\om^2 t + 3R \om_1^2\om t - y_1\om^4 t^2.
\end{eqnarray}
The initial position and velocity of the test body are 
$x_0$, $y_0$, $v_{0x}$, $v_{0y}$; the coordinates of the origin of the lab frame with respect to the Earth's centre are $x_1, \; y_1$, and the Earth-Sun distance is $R$.

The trajectory and the velocity 
can be obtained directly from Eq.~(\ref{9}) by integrating once or twice. Thus, the trajectory is given by a parametric 4th order curve. An interesting problem is to investigate  whether or not 
an experiment can be designed which would be based on this general solution.
\section{Conclusion}

To summarize,  it is proposed to test the validity of the modified Newtonian dynamics hypothesis in a laboratory based ``crucial experiment". The necessary and sufficient condition for entering the MOND regime  has been worked out for the most general type of motion. Particular cases of this condition (such as static, stationary, or non-stationary) suggest a variety of ways in which the experiment can be carried out in practice.  Performing such a test would require  a lot of precision.
 Fortunately, it is shown that the required high precision data can be borrowed from the existing sources such as the International Earth Rotation and Reference Systems Service (IERS). 

One interesting possibility is to experiment with a test body at rest. In this case, the familiar  techniques---such as gravitational wave detectors or torsion balances---can be useful. Both the time and the location of the experiment must satisfy rather strict requirements 
that have been derived here (the SHLEM effect). Despite obvious difficulties that they create, these restrictions carry some good news as they can help to improve extracting signal from  noise because the time of the SHLEM event can be predicted with high precision.
Another possibility is  based on the idea of cancellation between the centrifugal and the Coriolis inertial forces (the CCC set-up). In this case, although the condition  of the near-equinox synchronization should still hold, the location of the experiment is not restricted.

The author hopes that embarking on this challenging but important program by experimentalists will be a major step in  clarifying the status of the `MOND versus dark matter' dilemma.

Helpful discussions with L. Ignatieva and V.A.Kuzmin are gratefully acknowledged.


\begin{thebibliography}{99}
\bibitem{m0}M. Milgrom, Astrophys. J. {\bf270}, 365 (1983); {\em ibid.} {\bf270}, 371 (1983); {\em ibid.} {\bf270}, 384 (1983).
\bibitem{m} M. Milgrom, Ann. Phys. {\bf229}, 384 (1994); Phys. Lett. A {\bf253}, 273 (1999); arXiv: astro-ph/0510117;
J. D. Bekenstein,  Phys. Rev. D {\bf70}, 083509 (2004), {\em ibid.} D {\bf71}, 069901(E) (2005), arXiv: astro-ph/0403694; astro-ph/0412652; M. E. Soussa and R. P. Woodard, Phys. Lett. B {\bf578},  253 (2004);
Class. Quant. Grav. {\bf20}  2737 (2003); 
M. E. Soussa,
hep-th/0309150.
\bibitem{s}  
S. McGaugh, arXiv: astro-ph/9812328; Astrophys. J. {\bf611}, 26 (2004); arXiv: astro-ph/0606351; 
W. J. G. de Blok and S. S. McGaugh,
ÊÊarXiv: astro-ph/9805120;
A. Lue and G. D. Starkman Phys. Rev. Lett. {\bf92}, 131102 (2004); A. Slosar, A. Melchiorri, and J. I. Silk, Phys. Rev. D {\bf72}, 101301(R) (2005);  C. Skordis, 
D. F. Mota, P. G. Ferreira, and C. Boehm,
Phys. Rev. Lett. {\bf96}, 011301 (2006);  J. Bekenstein and J. Magueijo, Phys. Rev. D {\bf73}, 103513 (2006);  A. Knebe and B. K. Gibson, Mon. Not. R. Astron. Soc. {\bf347}, 1055 (2004); 
A. Knebe, arXiv: astro-ph/0509665; 
J. M. Romero and A. Zamora, Phys. Rev. D {\bf73},  027301 (2006); 
G. W. Angus, B. Famaey, H. S. Zhao, Mon. Not. R. Astron. Soc. {\bf371}, 138 (2006);
M. Kaplinghat and M. S. Turner, Astrophys. J. {\bf569},  L19 (2002);
M. Milgrom, arXiv: astro-ph/0110362; M. Milgrom and R. H. Sanders, arXiv:
astro-ph/0611494; S. Dodelson and M. Liguori, 
Phys. Rev. Lett. {\bf97}, 231301 (2006); V. Sahni and Y. Shtanov, arXiv: gr-qc/0606063; 
B.~Famaey and J. Binney, Mon. Not. R. Astron. Soc. {\bf363},  603 (2005); 
S. McGaugh's website http://www.astro.umd.edu/\~ssm/mond/ has a large bibliography on MOND.
\bibitem{rev}O. Bertolami and J. Paramos, arXiv: gr-qc/0611025; R. Scarpa, arXiv: astro-ph/0601478; 
 R. H. Sanders, arXiv: astro-ph/0601431;
H. S. Zhao,Ê arXiv: astro-ph/0508635.
\bibitem{icrs} http://aa.usno.navy.mil/faq/docs/ICRS\_doc.html\#REFS
\bibitem{itrs}http://itrf.ensg.ign.fr
\bibitem{ad}E. Adelberger (private communication) as discussed in \cite{h}.
\bibitem{h}C. Hagmann, arXiv: astro-ph/9905258.
\bibitem{adrev}E.G. Adelberger, B.R. Heckel, and A.E. Nelson,  Ann. Rev. Nucl. Part. Sci. {\bf53},  77 (2003). 
\bibitem{f}E. Fischbach and C. L. Talmadge, {\em The Search for Non-Newtonian Gravity} (Springer, New York, U.S.A., 1998); 
C. M. Will, Living Rev. Relativity {\bf9}, 3 (2006). 
\bibitem{a} {\em Allen's Astrophysical Quantities}, edited by A. N. Cox (AIP and Springer, Melville, 
1999) 4th edition.
\bibitem{fr}http://hpiers.obspm.fr /eop-pc
\bibitem{usno}http://maia.usno.navy.mil/bulletin-a.html
\bibitem{nist}http://www.nist.gov.
\bibitem{ign} A.~Yu.~Ignatiev (to be published).
\end{thebibliography}
\end{document}